\documentclass[11pt]{article}
\usepackage{epsfig} 
\newcommand{\Dslash}{D \! \! \! \! /}

\newcommand{\half}{\mbox{\small{$\frac{1}{2}$}}} 
\newcommand{\MSbar}{\overline{\mbox{MS}}} 
 
\newcommand{\Nf}{N_{\!f}}
\newcommand{\NF}{N_{\!F}}
\newcommand{\NA}{N_{\!A}}

\begin{document}
\title{Three loop $\MSbar$ anomalous dimension for renormalizable gauge 
invariant non-local gluon mass operator in QCD}
\author{F.R. Ford \& J.A. Gracey, \\ Theoretical Physics Division, \\ 
Department of Mathematical Sciences, \\ University of Liverpool, \\ P.O. Box 
147, \\ Liverpool, \\ L69 3BX, \\ United Kingdom.} 
\date{} 
\maketitle 
\vspace{5cm} 
\noindent 
{\bf Abstract.} The three loop anomalous dimension for the gauge invariant,
renormalizable, non-local mass operator for a gluon is computed in the 
$\MSbar$ scheme. In addition the anomalous dimensions of the associated 
localizing ghost fields are also deduced at the same order and it is shown that
the three loop QCD $\beta$-function correctly emerges from the gluon localizing
ghost vertex renormalization.  

\vspace{-17.5cm}
\hspace{13.5cm}
{\bf LTH 826}

\newpage 

In non-abelian gauge theories the vector bosons responsible for carrying the
quanta of force are regarded as massless particles unless there is a 
spontaneous symmetry breaking. Expressed another way there is no gauge 
invariant local mass operator for gluons in quantum chromodynamics (QCD). 
Whilst it is possible to have BRST invariant masses, such as that which occurs
in the Curci-Ferrari model, \cite{1}, the inclusion of such local mass 
operators all suffer from the disadvantage of leading to non-unitary theories, 
\cite{2,3}. Hence they have no predictive power in relation to $S$-matrix 
elements. By contrast, there has been an explosion of interest in recent years 
in studying the infrared dynamics of Yang-Mills theories in the infrared limit 
in the Landau gauge using lattice techniques, Dyson-Schwinger equation methods 
and other more formal approaches. One of the main quantities which is analysed 
is the gluon propagator and it is widely acknowledged that it does not satisfy 
the usual perturbative form of a massless propagator of an unconfined field. 
Instead it is generally fair to say that the gluon propagator, as measured on 
the lattice and other methods, has a behaviour which is not inconsistent with 
the gluon having an effective mass of some sort. Whether this effective mass is
due to screening, dynamically generated, derived from say Gribov issues, due to
vortex condensation or another mechanism has not yet been definitively 
answered. However, if it is to be explained theoretically then one is forced 
into studying extensions of the Yang-Mills or QCD Lagrangians which have a 
concrete gluon mass term of some sort or one where a mass operator condenses. 
Clearly to do this in a gauge invariant way would appear impossible as the 
obvious mass operator, $\half ( A^a_\mu )^2$, breaks gauge symmetry despite 
being renormalizable, \cite{1}, where $A^a_\mu$ is the gluon field. However, if
one sacrifices the restriction to {\em local} operators then it is possible to 
have several gauge invariant gluon mass terms. In essence there are two types.

The first, originally introduced in \cite{4} in three dimensions, has been
examined in four dimensions in \cite{5,6} where it was shown to be 
renormalizable. Indeed its two loop $\MSbar$ anomalous dimension was computed 
in \cite{6} and shown to be independent of the gauge fixing parameter of a 
linear covariant gauge. The key to demonstrating renormalizability and allowing
one to calculate in a systematic way was the fact that the Lagrangian involving
the operator itself could be written in terms of local fields additional to the
usual gluon, quark and Faddeev-Popov ghost fields. These extra (infrared) 
fields do not affect the usual ultraviolet properties of the original 
non-abelian gauge theory, \cite{5,6}. Therefore, for example, the 
$\beta$-function of \cite{7} is unchanged. The other type of mass operator is 
in effect the Stueckelberg term but written as 
$\stackrel{\mbox{\begin{small}min\end{small}}}
{\mbox{\begin{tiny}$\{U\}$\end{tiny}}} \int d^4x \, ( A^{a \, U}_\mu )^2$ 
where $U$ is an element of the gauge group. It has been studied in the massive 
gauge invariant model in \cite{8,9} and is central to a vortex interpretation 
of confinement. Although also being non-local it suffers from the calculational
drawback of being non-renormalizable. Though its one loop anomalous dimension 
was computed in \cite{10} for arbitrary linear covariant gauge and shown to be 
independent of the gauge parameter. Part of that calculation rested on the fact
that the massive gauge invariant operator  
$\stackrel{\mbox{\begin{small}min\end{small}}}
{\mbox{\begin{tiny}$\{U\}$\end{tiny}}} \int d^4x \, ( A^{a \, U}_\mu )^2$ has 
the renormalizable non-local mass operator of \cite{5,6} as its first term in a
gluon leg expansion of the operator in terms of gauge invariant operators,
\cite{11}. Therefore the localization of the previous non-local operator into 
the original Yang-Mills fields plus localizing ghost fields provided a useful 
calculational shortcut. From another point of view the non-local operator 
$\stackrel{\mbox{\begin{small}min\end{small}}}
{\mbox{\begin{tiny}$\{U\}$\end{tiny}}} \int d^4x \, ( A^{a \, U}_\mu )^2$ can 
be viewed as a method of gauge fixing QCD in a more concrete fashion as noted
in \cite{11,12,13,14}. This is because that gauge fixing operator is gauge 
invariant and thus avoids the Gribov problem, \cite{15}, which plagues the more
widely used Landau gauge in the present intense activity into the infrared 
structure. 

From a theoretical point of view one would ultimately like to have a Lagrangian
based method of studying effective gluon mass which emerges in the current
picture and which is renormalizable. Moreover, as performing calculations is
essential to understanding such low energy problems, we focus here on providing
the anomalous dimensions of the non-local mass operator of \cite{5,6} to 
{\em three} loops in the $\MSbar$ scheme. This is far from being a trivial 
exercise which is due in part to the presence of the additional fields but also
because of the generation of a set of quartic interactions. As was shown in 
\cite{5,6} these are essential to preserving multiplicative renormalizability. 
Therefore, we also report on the renormalization of the fields themselves at 
three loops. Indeed as an example of where such three loop results are 
necessary we note that in \cite{16,17} the problem of the dynamical generation 
of a gluon mass was studied in the Landau gauge based on the local operator 
$\half ( A^a_\mu )^2$. Briefly, the two loop effective potential for this 
operator was computed for $\Nf$ massless quarks using the local composite 
operator (LCO) formalism, \cite{18,19}. Knowledge of this potential allows one 
to show that the energetically favoured vacuum is one where the operator 
condenses and therefore dynamically generates a gluon mass. One peculiar 
feature of the LCO formalism, however, is that to have the full two loop 
potential one needs the operator's anomalous dimension at {\em three} loops, 
\cite{16,17,18,19}. Whilst the results successfully demonstrated operator 
condensation which was stable to loop corrections, \cite{16,17}, it suffers 
from one obvious drawback and that is that the calculation was restricted to a 
specific gauge. It would be more appropriate to study the extended operator 
considered here since it is gauge invariant. Indeed this is one of our 
motivations for this article. However, as will be evident from what we present,
we believe the determination of this three loop anomalous dimension for the 
non-local operator is sufficiently interesting in its own right to present it 
separate from an LCO computation.

We begin by recalling the full form of the Lagrangian of \cite{5,6}. It is 
\begin{eqnarray} 
L &=& -~ \frac{1}{4} G_{\mu\nu}^a G^{a \, \mu\nu} ~-~ \frac{1}{2\alpha} 
(\partial^\mu A^a_\mu)^2 ~-~ \bar{c}^a \partial^\mu D_\mu c^a ~+~ 
i \bar{\psi}^{iI} \Dslash \psi^{iI} \nonumber \\ 
&& +~ \frac{1}{4} \left( \bar{B}^a_{\mu\nu} D^{ab}_\sigma D^{bc \, \sigma}
B^{c \, \mu\nu} ~-~ \bar{H}^a_{\mu\nu} D^{ab}_\sigma D^{bc \, \sigma}
H^{c \, \mu\nu} \right) ~+~ \frac{im}{4} \left( B^a_{\mu\nu} 
- \bar{B}^a_{\mu\nu} \right) G^{a\,\mu\nu} \nonumber \\
&& +~ \frac{1}{16} \lambda^{abcd} 
\left( \bar{B}^a_{\mu\nu} B^{b\,\mu\nu} - \bar{H}^a_{\mu\nu} H^{b\,\mu\nu}
\right) \left( \bar{B}^c_{\sigma\rho} B^{d\,\sigma\rho} 
- \bar{H}^c_{\sigma\rho} H^{d\,\sigma\rho} \right) 
\label{lag}
\end{eqnarray}  
where $\alpha$ is the linear covariant gauge fixing parameter, $c^a$ is the 
Faddeev-Popov ghost, $\psi^{iI}$ is the (massless) quark, $B^a_{\mu\nu}$ and 
$H^a_{\mu\nu}$ are the localizing ghosts where the latter are anticommuting and
$m$ is the gluon mass. For completeness the index ranges are 
$1$~$\leq$~$a$~$\leq$~$\NA$, $1$~$\leq$~$I$~$\leq$~$\NF$ and 
$1$~$\leq$~$i$~$\leq$~$N_f$ where $N_F$ and $N_A$ are the respective dimensions
of the fundamental and adjoint representations and $\Nf$ is the number of 
quarks. The covariant derivative, involving the coupling constant $g$, is 
denoted by $D^{ab}_\mu$ and $G^a_{\mu\nu}$ is the field strength. The 
quantities $\lambda^{abcd}$ are the quartic couplings necessary for 
multiplicative renormalizability and satisfy the symmetry properties  
\begin{equation}
\lambda^{abcd} ~=~ \lambda^{bacd} ~=~ \lambda^{abdc} ~=~ \lambda^{cdab} ~.
\label{lamsym}
\end{equation}
They are not to be confused with the specific rank $4$ invariant tensors, such
as the totally symmetric tensor $d_F^{abcd}$ of \cite{20}, which can be built 
from the structure functions, $f^{abc}$, or the colour group generators, $T^a$.
In addition, since the Lagrangian is colour symmetric the quartic couplings 
satisfy a Jacobi style identity, \cite{5,6}, which is
\begin{equation}
f^{apq} \lambda^{pbcd} ~+~ f^{bpq} \lambda^{apcd} ~+~ f^{cpq} 
\lambda^{abpd} ~+~ f^{cpq} \lambda^{abcp} ~=~ 0 ~.
\label{lamjac}
\end{equation}
In (\ref{lag}) we have ignored the masses of the $\{ B^a_{\mu\nu}, 
\bar{B}^a_{\mu\nu}, H^a_{\mu\nu}, \bar{H}^a_{\mu\nu} \}$ sector since they will
play no role in the present calculation. Finally, we note that (\ref{lag}) is 
the localized version of the Lagrangian with the explicit non-local mass 
operator, \cite{5},
\begin{equation} 
L ~=~ -~ \frac{1}{4} G_{\mu\nu}^a G^{a \, \mu\nu} ~-~ \frac{1}{2\alpha} 
(\partial^\mu A^a_\mu)^2 ~-~ \bar{c}^a \partial^\mu D_\mu c^a ~+~ 
i \bar{\psi}^{iI} \Dslash \psi^{iI} ~-~ \frac{m^2}{4} G^a_{\mu\nu} 
\left( \frac{1}{D^2} \right)^{ab} G^{b\,\mu\nu} 
\end{equation} 
where $D^2$ is the square of the covariant derivative. 

Clearly with the additional ghost fields and coupling one has to ensure that
the gluon, ghost and quark anomalous dimensions as well as the usual
$\beta$-function remain independent of $\lambda^{abcd}$. This has been verified
at two loops in \cite{5} and \cite{6}. Therefore, here we will compute the 
former anomalous dimensions to three loops as well as those for $B^a_{\mu\nu}$ 
and $H^a_{\mu\nu}$. The latter will be $\lambda^{abcd}$-dependent. To deduce 
the anomalous dimension of the mass $m$ or equivalently the mass operator 
anomalous dimension we will renormalize the dimension three gauge invariant 
operator ${\cal O}$ where 
\begin{equation}
{\cal O} ~=~ \frac{1}{4} \left( B^a_{\mu\nu} - \bar{B}^{a\,\mu\nu} \right) 
G^{a\,\mu\nu} 
\end{equation}
by inserting it into a gluon $B^a_{\mu\nu}$ two point function. The advantage 
of this approach is that one can split the free and interaction Lagrangian in 
such a way that the operator is in the latter and not the former. If it were 
included in the free part then we would have the huge (and unnecessary)
computational task of calculating with massive propagators which would require
the inclusion of the masses of $\{ B^a_{\mu\nu}, \bar{B}^a_{\mu\nu}, 
H^a_{\mu\nu}, \bar{H}^a_{\mu\nu} \}$. (The explicit mass terms are given in
\cite{5,6}.) This would be an intractable proposition. Instead treating the 
operator as an insertion means that all fields remain massless and one also 
avoids the mixing of masses which occurs in the full quadratic sector of such a
Lagrangian split, aside from the additional complications from the 
$B^a_{\mu\nu}$ and $H^a_{\mu\nu}$ masses. More crucially with massless fields 
one can employ the {\sc Mincer} algorithm, \cite{21}, which has been encoded, 
\cite{22}, in the symbolic manipulation language {\sc Form}, \cite{23}. The 
{\sc Mincer} procedure applies to massless three loop $2$-point functions, 
\cite{21}, and performs the computation in dimensional regularization in 
$d$~$=$~$4$~$-$~$2\epsilon$ dimensions where $\epsilon$ is the regularizing 
parameter. Such a high loop order calculation can clearly only be performed via
automatic Feynman diagram techniques. In such an approach the extraction of the
operator anomalous dimension is relegated to the evaluation of the divergent 
part of a $2$-point function derived from the parent $3$-point one, 
$\langle A^a_{\mu} {\cal O} B^b_{\nu\sigma} \rangle$, where the external 
momentum of the $B^b_{\nu\sigma}$ field is nullified. Such a process for this 
Green's function is infrared safe since no infrared divergent factors such as 
$1/(k^2)^2$ arise in a Feynman integral where $k$ is an internal momentum. 

{\begin{table}[ht] 
\begin{center} 
\begin{tabular}{|c||c|c|c|c|c|} 
\hline 
Green's function & One loop & Two loop & Three loop & Total \\ 
\hline 
$ A^a_\mu \, A^b_\nu$ & $~5$ & $~52$ & $~~1279$ & $~~1336$ \\ 
$ c^a \, {\bar c}^b$ & $~1$ & $~~8$ & $~~~152$ & $~~~~161$ \\ 
$ \psi^{iI} \, {\bar \psi}^{jJ}$ & $~1$ & $~~8$ & $~~~152$ & $~~~~161$ \\ 
$ B^a_{\mu\nu} \, {\bar B}^b_{\sigma\rho}$ & $~1$ & $~20$ & $~~~464$ & 
$~~~~485$ \\
$ H^a_{\mu\nu} \, {\bar H}^b_{\sigma\rho}$ & $~1$ & $~20$ & $~~~464$ & 
$~~~~485$ \\
$ A^a_\mu \, {\bar B}^b_{\nu\sigma} \, B^c_{\rho\phi}$ & $~7$ & $166$ & 
$~~5827$ & $~~6000$ \\ 
$ A^a_\mu \, {\cal O} \, {\bar B}^b_{\nu\sigma} $ & $~5$ & $131$ & 
$~~6917$ & $~~7053$ \\ 
\hline 
Total & $21$ & $405$ & $15255$ & $15681$ \\  
\hline 
\end{tabular} 
\end{center} 
\begin{center} 
{Table 1. Number of Feynman diagrams for each Green's function.} 
\end{center} 
\end{table}}  

In the final part of this setup description we note that we have to be careful 
in ensuring the correctness of the final expression. Since the operator 
insertion is in a Green's function involving a localizing ghost, we require a 
strong check on the $B^a_{\mu\nu}$ renormalization constants. To ensure this we
have also performed the three loop $\MSbar$ renormalization of the 
$A^a_\mu \bar{B}^b_{\nu\sigma} B^c_{\rho\phi}$ vertex itself. As in the 
original QCD Lagrangian, this Green's function will produce the three loop 
$\MSbar$ $\beta$-function of the gauge coupling, \cite{7,24,25}. Again as this 
is a $3$-point function we nullify the external momentum of the 
$B^c_{\rho\phi}$ field relegating it to a $2$-point function whence it can be 
determined by the {\sc Mincer} algorithm. In Table $1$, we have listed the 
number of Feynman diagrams computed for the present article. Those for the 
gluon and ghost exceed the numbers for the corresponding original QCD 
calculations due to the presence of the localizing fields. The numbers of 
graphs in Table $1$ are deduced from the {\sc Qgraf} package, \cite{26}, which 
is the starting point for each of the Green's functions. The {\sc Qgraf} 
routine generates the Feynman diagrams electronically and these are then 
converted to {\sc Form} input notation prior to the application of the 
{\sc Mincer} algorithm. One additional complication is the non-trivial task of 
extending the {\sc Form} group theory module to handle the group theory 
associated with the $\lambda^{abcd}$ couplings subject to the symmetry and 
Jacobi properties of (\ref{lamsym}) and (\ref{lamjac}). In addition we have 
also used the property noted in \cite{6} that
\begin{equation}
\lambda^{acde} \lambda^{bcde} ~=~ \frac{1}{\NA} \delta^{ab} \lambda^{cdpq}
\lambda^{cdpq} ~~,~~ 
\lambda^{acde} \lambda^{bdce} ~=~ \frac{1}{\NA} \delta^{ab} \lambda^{cdpq}
\lambda^{cpdq} 
\end{equation} 
and the analogous extension to the products of three $\lambda^{abcd}$-tensors 
with two free indices, which follow from the fact that there is only one rank 
two isotropic tensor in a classical Lie group. Finally, we note that the 
propagators of the (massless) localizing ghosts are given in \cite{20}.  

We now record our main results at three loops. First, we define the
renormalization constants for the relevant fields and the operator as
\begin{equation}
B^{a \, \mu\nu}_{\mbox{\footnotesize{o}}} ~=~ \sqrt{Z_B} \, 
B^{a \, \mu\nu} ~~,~~ 
H^{a \, \mu\nu}_{\mbox{\footnotesize{o}}} ~=~ \sqrt{Z_H} \, 
H^{a \, \mu\nu} ~~,~~ 
{\cal O}_{\mbox{\footnotesize{o}}} ~=~ Z_{\cal O} {\cal O} 
\label{Zdef}
\end{equation}
where the subscript ${}_{\mbox{\footnotesize{o}}}$ denotes the bare quantity.
Then the respective anomalous dimensions are
\begin{equation} 
\gamma_B(a,\lambda,\alpha) ~=~ \mu \frac{d~}{d\mu} \ln Z_B ~~~,~~~  
\gamma_H(a,\lambda,\alpha) ~=~ \mu \frac{d~}{d\mu} \ln Z_H ~~~,~~~  
\gamma_{\cal O}(a,\lambda) ~=~ \mu \frac{d~}{d\mu} \ln Z_{\cal O} 
\end{equation}
where we note
\begin{equation}
\mu \frac{d~}{d\mu} ~=~ \beta(a) \frac{\partial~}{\partial a} ~+~ 
\beta^{pqrs}_\lambda(a,\lambda) \frac{\partial~}{\partial \lambda^{pqrs}} ~+~
\alpha \gamma_\alpha(a,\alpha) \frac{\partial~}{\partial\alpha} 
\end{equation}
with $\beta(a)$ the $\beta$-function of the gauge coupling 
$a$~$=$~$g^2/(16\pi^2)$ and $\gamma_\alpha(a,\alpha)$ is the anomalous 
dimension of the linear covariant gauge fixing parameter. We use the 
conventions of \cite{27} for this and note that 
$\gamma_\alpha(a,\alpha)$~$=$~$-$~$\gamma_A(a,\alpha)$ with the latter defined 
to be the gluon anomalous dimension. In the term involving the 
$\beta$-function of the quartic couplings, $\beta^{abcd}(a,\lambda)$, it is 
understood that the differentiation respects the symmetries of the 
$\lambda^{abcd}$-tensor couplings, (\ref{lamsym}). Given this we find the 
$\MSbar$ expression
\begin{eqnarray}
\gamma_B(a,\lambda,\alpha) &=& \gamma_H(a,\lambda,\alpha) \nonumber \\
&=& ( \alpha - 3 ) C_A ~+~ \left[ \left( \frac{1}{4}\alpha^2 + 2\alpha 
- \frac{61}{6} \right) C_A^2 + \frac{10}{3} T_F \Nf C_A \right] a^2 ~+~ 
\frac{1}{128\NA} \lambda^{abcd} \lambda^{acbd} \nonumber \\ 
&& +~ \left[ \left( \frac{5}{16} \alpha^3 + \frac{39}{32}\alpha^2 
+ \frac{271}{32} \alpha - \frac{18193}{432} + \left( \frac{3}{8} \alpha^2
- \frac{27}{8} \right) \zeta(3) \right) C_A^3 \right. \nonumber \\
&& ~~~~~ \left. +~ \left( \frac{5}{54} + 48 \zeta(3) - \frac{17}{4} \alpha
\right) T_F \Nf C_A^2 ~+~ \left( 45 - 48 \zeta(3) \right) T_F \Nf C_F C_A
\right. \nonumber \\ 
&& ~~~~~ \left. +~ \frac{140}{27} T_F^2 \Nf^2 C_A \right] a^3 ~+~
\left[ \frac{3}{8} \zeta(3) - \frac{13}{64} \right] \frac{C_A}{\NA}
f_4^{abcd} \lambda^{acbd} a^2 \nonumber \\
&& +~ \frac{1}{\NA} \left[ \frac{13}{16} - \frac{3}{2} \zeta(3) \right]  
f_4^{abcd} f_4^{apcq} \lambda^{bpdq} a^2 ~+~ \frac{5C_A}{64\NA} \lambda^{abcd} 
\lambda^{acbd} a \nonumber \\
&& -~ \frac{1}{2048\NA} \left[ 3 \lambda^{abcd} \lambda^{acpq} \lambda^{bpdq} 
+ \lambda^{abcd} \lambda^{apcq} \lambda^{bqdp} \right] ~+~ O(a^4;\lambda^4) 
\label{anomb}
\end{eqnarray}
where $\zeta(n)$ is the Riemann zeta function and a factor of $1/(4\pi)$, which
derives from the loop integral measure, has formally been absorbed into each 
$\lambda^{abcd}$ to simplify the presentation. The result (\ref{anomb}) 
explicitly verifies the equality of the Slavnov-Taylor identity of \cite{5,6} 
to three loops. The group Casimirs are defined by $\mbox{Tr} \left( T^a T^b 
\right)$~$=$~$T_F\delta^{ab}$, $T^aT^a$~$=$~$C_F$ and $f^{acd} 
f^{bcd}$~$=$~$C_A \delta^{ab}$. We have introduced the shorthand notation for 
the contraction of two structure functions
\begin{equation}
f_4^{abcd} ~=~ f^{eab} f^{ecd}
\end{equation}
and defined the order symbol, $O(a^4;\lambda^4)$, to correspond to the four 
loop corrections. Moreover, with these values the three loop $\MSbar$ QCD 
$\beta$-function of \cite{7,24,25} correctly emerges as $\lambda^{abcd}$ and 
$\alpha$ independent from the $A^a_\mu\bar{B}^b_{\nu\sigma} B^c_{\rho\phi}$ 
vertex. Given this we find the three loop correction to the gluon mass operator
is
\begin{eqnarray}
\gamma_{\cal O}(a,\lambda) &=& \left[ \frac{11}{6} C_A - \frac{2}{3} T_F \Nf 
\right] a ~+~ \left[ \frac{77}{24} C_A^2 - \frac{2}{3} T_F \Nf C_A 
- 2 T_F \Nf C_F \right] a^2 \nonumber \\
&& -~ \frac{1}{16\NA} f_4^{abcd} \lambda^{acbd} a ~-~ \frac{1}{256\NA} 
\lambda^{abcd} \lambda^{acbd} \nonumber \\  
&& +~ \left[ \frac{361}{32} C_A^3 - \frac{211}{36} T_F \Nf C_A^2 
- \frac{97}{18} T_F \Nf C_F C_A + T_F \Nf C_F^2 \right. \nonumber \\
&& ~~~~~ \left. +~ \frac{5}{9} T_F^2 \Nf^2 C_A + \frac{22}{9} T_F^2 \Nf^2 C_F 
\right] a^3 ~+~ \frac{19}{32\NA} f_4^{abcd} f_4^{apcq} \lambda^{bpdq} a^2 
\nonumber \\
&& -~ \frac{1}{\NA} \left[ \frac{1}{144} T_F \Nf + \frac{857}{1152} C_A
\right] f_4^{abcd} \lambda^{acbd} a^2 ~-~ \frac{19C_A}{512\NA} \lambda^{abcd} 
\lambda^{acbd} a \nonumber \\
&& +~ \frac{1}{\NA} \left[ \frac{31}{768} f_4^{abcd} \lambda^{apcq} 
\lambda^{bdpq} + \frac{9}{512} f_4^{abcd} \lambda^{apbq} \lambda^{cpdq} 
- \frac{25}{768} f_4^{abcd} \lambda^{acpq} \lambda^{bpdq} \right] a
\nonumber \\
&& +~ \frac{1}{4096\NA} \left[ 3 \lambda^{abcd} \lambda^{acpq} \lambda^{bpdq} 
+ \lambda^{abcd} \lambda^{apcq} \lambda^{bqdp} \right] ~+~ O(a^4;\lambda^4) 
\label{anomop}
\end{eqnarray} 
which is clearly $\alpha$ independent as expected on general grounds but which
in fact provides a non-trivial check on our computation. It is worth stressing
that the emergence of the correct $\lambda^{abcd}$ independent 
$\beta$-function and the gauge parameter independent 
$\gamma_{\cal O}(a,\lambda)$ is a non-trivial check on the implementation of 
the symmetry properties of $\lambda^{abcd}$ in the {\sc Form} group theory
module.

One additional calculational detail is worth noting and that is that 
$\lambda^{abcd}$ itself undergoes a renormalization within the three loop 
calculations. Its {\em one} loop $\beta$-function was given in \cite{6} as
\begin{eqnarray}
\beta_\lambda^{abcd}(a,\lambda) &=& \frac{1}{2} (d-4) \lambda^{abcd} ~+~ 
\frac{1}{8} \left[ \lambda^{abpq} \lambda^{cqdp} + \lambda^{apbq} 
\lambda^{cdpq} + \lambda^{apcq} \lambda^{bpdq} + \lambda^{apdq} \lambda^{bpcq} 
\right] \nonumber \\
&& -~ 6 C_A \lambda^{abcd} a ~-~ 12 C_A f_4^{abcd} a^2 ~+~ 48 f_4^{apbq}
f_4^{cpdq} a^2 ~+~ O(a^3;\lambda^3) ~. 
\label{betalam}
\end{eqnarray} 
Since we are going one loop beyond \cite{6}, it might have been expected that 
the two loop $\MSbar$ correction of (\ref{betalam}) was needed. However, for 
the operator renormalization the first place $\lambda^{abcd}$ occurs is at two 
loops. Therefore, one only needs its one loop renormalization. Equally for the 
$B^a_{\mu\nu}$ and $H^a_{\mu\nu}$ anomalous dimensions $\lambda^{abcd}$ first 
appears at two loops and again only its one loop renormalization is necessary 
to deduce the fully renormalized three loop $2$-point function. Here this is 
because the one loop graph involving $\lambda^{abcd}$ which contributes to 
either $2$-point function results in a snail graph which clearly is zero for 
the massless fields we consider. In other words it would only contribute to the
renormalization of the $B^a_{\mu\nu}$ or $H^a_{\mu\nu}$ mass renormalization. 
Such a property of the $\lambda^{abcd}$-structure of the anomalous dimensions 
in fact prevents us from having to extend the $\lambda^{abcd}$ renormalization 
to two loops by renormalizing massless $4$-point functions which have non-safe 
nullifiable external momenta and hence not accessible to the {\sc Mincer} 
algorithm. Moreover, it is worth noting that this is the first use of 
(\ref{betalam}) within a computation and the overall consistency of our three
loop renormalization is a non-trivial check on its correctness.  

As a final check on our anomalous dimensions, we note that in the original
renormalization constants we have been careful to check that the triple and
double poles in $\epsilon$ are correctly predicted from the known one and two
loop structure. For the current Lagrangian, (\ref{lag}), this has an additional
feature and that is that one has to take into account two coupling constants,
$a$ and $\lambda^{abcd}$. To aid the interested reader in this respect, we
provide the explicit three loop $\MSbar$ renormalization constant for
${\cal O}$ whence (\ref{anomop}) was deduced. It is\footnote{We note that
attached to the version of this article which appears on the arXiv there is a 
{\sc Form} file which contains the results (\ref{anomb}), (\ref{anomop}) and
(\ref{zop}).}
\begin{eqnarray}
Z_{\cal O} &=& 1 ~+~ \left[ \frac{2}{3} T_F \Nf - \frac{11}{6} C_A \right]
\frac{a}{\epsilon} ~+~ \left[ \frac{121}{24} C_A^2 + \frac{2}{3} T_F^2 \Nf^2
- \frac{11}{3} T_F \Nf C_A \right] \frac{a^2}{\epsilon^2} \nonumber \\ 
&& +~ \left[ \left( \frac{1}{3} T_F \Nf C_A - \frac{77}{48} C_A^2 
+ T_F \Nf C_F \right) a^2 + \frac{1}{512\NA} \lambda^{abcd} \lambda^{acbd}
+ \frac{1}{32\NA} f_4^{abcd} \lambda^{acbd} \right] \frac{1}{\epsilon}  
\nonumber \\ 
&& +~ \left[ \frac{605}{36} T_F \Nf C_A^2 - \frac{6655}{432} C_A^3 
- \frac{55}{9} T_F^2 \Nf^2 C_A + \frac{20}{27} T_F^3 \Nf^3 \right] 
\frac{a^3}{\epsilon^3} \nonumber \\ 
&& +~ \left[ \left( \frac{3989}{288} C_A^3 - \frac{757}{72} T_F \Nf C_A^2 
- \frac{121}{18} T_F \Nf C_A C_F + 2 T_F^2 \Nf^2 C_A + \frac{22}{9} T_F^2 \Nf^2
C_F \right) a^3 \right. \nonumber \\
&& ~~~~~ \left. +~ \frac{1}{6144\NA} \left( 
3 \lambda^{abcd} \lambda^{acpq} \lambda^{bpdq} 
+ \lambda^{abcd} \lambda^{apcq} \lambda^{bqdp} \right) \right. \nonumber \\
&& ~~~~~ \left. +~ \frac{1}{\NA} \left( 
\frac{1}{384} f_4^{abcd} \lambda^{acpq} \lambda^{bdpq} 
+ \frac{1}{256} f_4^{abcd} \lambda^{apbq} \lambda^{cpdq} 
+ \frac{1}{384} f_4^{abcd} \lambda^{apcq} \lambda^{bpdq} \right) a \right.
\nonumber \\ 
&& ~~~~~ \left. -~ \frac{C_A}{\NA} \left( 
\frac{1}{384} \lambda^{abcd} \lambda^{abcd} 
+ \frac{31}{3072} \lambda^{abcd} \lambda^{acbd} \right) a 
+ \frac{T_F\Nf}{768\NA} \lambda^{abcd} \lambda^{acbd} a \right. \nonumber \\ 
&& ~~~~~ \left. +~ \frac{1}{16\NA} f_4^{abcd} f_4^{apcq} \lambda^{bqdp} a^2 
- \frac{41C_A}{288\NA} f_4^{abcd} \lambda^{acbd} a^2 
+ \frac{5T_F\Nf}{144\NA} f_4^{abcd} \lambda^{acbd} a^2 \right]
\frac{1}{\epsilon^2} \nonumber \\ 
&& +~ \left[ \left( \frac{211}{108} T_F \Nf C_A^2 - \frac{361}{96} C_A^3 
+ \frac{97}{54} T_F \Nf C_A C_F - \frac{5}{27} T_F^2 \Nf^2 C_A 
- \frac{22}{27} T_F^2 \Nf^2 C_F \right. \right. \nonumber \\
&& ~~~~~ \left. \left. -~ \frac{1}{3} T_F \Nf C_F^2 \right) a^3 
- \frac{1}{12288\NA} \left( 3 \lambda^{abcd} \lambda^{acpq} \lambda^{bpdq} 
+ \lambda^{abcd} \lambda^{apcq} \lambda^{bqdp} \right) \right. \nonumber \\
&& ~~~~~ \left. +~ \frac{1}{\NA} \left( 
\frac{5}{128} f_4^{abcd} \lambda^{acpq} \lambda^{bdpq} 
- \frac{31}{384} f_4^{abcd} \lambda^{acpq} \lambda^{bpdq} 
\right. \right. \nonumber \\
&& ~~~~~~~~~~~~~~~~ \left. \left. +~ \frac{17}{512} f_4^{abcd} \lambda^{apbq} 
\lambda^{cpdq}
+ \frac{5}{128} f_4^{abcd} \lambda^{apcq} \lambda^{bpdq} \right) a \right.
\nonumber \\ 
&& ~~~~~ \left. +~ \frac{C_A}{\NA} \left( 
\frac{79}{1536} \lambda^{abcd} \lambda^{acbd} 
- \frac{5}{128} \lambda^{abcd} \lambda^{abcd} \right) a 
- \frac{19}{96\NA} f_4^{abcd} f_4^{apcq} \lambda^{bqdp} a^2 \right. 
\nonumber \\ 
&& ~~~~~ \left. +~ \frac{515C_A}{3456\NA} f_4^{abcd} \lambda^{acbd} a^2 
+ \frac{T_F\Nf}{432\NA} f_4^{abcd} \lambda^{acbd} a^2 \right]
\frac{1}{\epsilon} ~+~ O(a^4;\lambda^4) ~.  
\label{zop}
\end{eqnarray}
We note that the extraction of all our renormalization constants followed the
procedure discussed in \cite{28} for automatic Feynman diagram computations
where all the Green's functions are first determined as a function of the bare 
quantities, such as the coupling constant. The renormalized versions are then 
introduced by the standard rescaling and the remaining divergences in 
$\epsilon$ are absorbed into the appropriate renormalization constant 
associated with that particular Green's function. This procedure is 
straightforward to implement within the {\sc Form} routines.

To conclude we have provided the full three loop $\MSbar$ renormalization of 
all quantities in the localizing Lagrangian, (\ref{lag}), for a renormalizable
gauge invariant non-local mass operator, except the quartic couplings. Whilst
this has been a cumbersome task it will actually play a crucial role when one
considers the corresponding renormalization of the full operator of 
$\stackrel{\mbox{\begin{small}min\end{small}}}
{\mbox{\begin{tiny}$\{U\}$\end{tiny}}} \int d^4x \, ( A^{a \, U}_\mu )^2$ 
beyond the one loop result of \cite{10}. Although this is clearly 
non-renormalizable, it is not inconceivable that one could localize the 
operator order by order in perturbation theory with a finite set of localizing
fields at each order. Indeed the one loop result of \cite{10} proceeded under
this assumption. Whilst the additional couplings are absent at that level it
would be interesting to see the structure of the gauge parameter independent
anomalous dimension which emerges and to study the role of extra quartic
couplings play in any renormalization group evolution. Moreover, given the 
successful extraction of the {\em three} loop anomalous dimension, it now also 
opens up the possibility of computing the two loop effective potential of this 
gauge invariant operator to study its condensation properties.

\vspace{1cm}
\noindent
{\bf Acknowledgements.} 
The authors thank Prof. S.P. Sorella and Dr D. Dudal for useful discussions
and F.R. Ford thanks the University of Liverpool for a Research Studentship.


\begin{thebibliography}{99} 
\bibitem{1} G. Curci \& R. Ferrari, Nuovo Cim. {\bf A32} (1976), 151. 
\bibitem{2} G. Curci \& R. Ferrari, Nuovo Cim. {\bf A35} (1976), 1; 
G. Curci \& R. Ferrari, Nuovo Cim. {\bf A35} (1976), 273; 
Nuovo Cim. {\bf A47} (1978), 555. 
\bibitem{3} I. Ojima, Z. Phys. {\bf C13} (1982), 173.  
\bibitem{4} R. Jackiw \& S.Y. Pi, Phys. Lett. {\bf B403} (1997), 297. 
\bibitem{5} M.A.L. Capri, D. Dudal, J.A. Gracey, V.E.R. Lemes, R.F. Sobreiro, 
S.P. Sorella \& H. Verschelde, Phys. Rev. {\bf D72} (2005), 105016.
\bibitem{6} M.A.L. Capri, D. Dudal, J.A. Gracey, V.E.R. Lemes, R.F. Sobreiro, 
S.P. Sorella \& H. Verschelde, Phys. Rev. {\bf D74} (2006), 045008.
\bibitem{7} D.J. Gross \& F.J. Wilczek, Phys. Rev. Lett. {\bf 30}
(1973), 1343; 
H.D. Politzer, Phys. Rev. Lett. {\bf 30} (1973), 1346.
\bibitem{8} J.M. Cornwall, Phys. Rev. {\bf D26} (1982), 1453. 
\bibitem{9} J.M. Cornwall \& A. Soni, Phys. Lett. {\bf B120} (1983), 431. 
\bibitem{10} J.A. Gracey, Phys. Lett. {\bf B651} (2007), 253. 
\bibitem{11} D. Zwanziger, Nucl. Phys. {\bf B345} (1990), 461. 
\bibitem{12} M.A. Semenov-Tian-Shanskii \& V.A. Franke, Zap. Nauchn. Semin.
LOMI {\bf 120} (1982), 159; J. Sov. Math. {\bf 34} (1986), 1999.
\bibitem{13} G. Dell'Antonio \& D. Zwanziger, Commun. Math. Phys. {\bf 138}
(1991), 291. 
\bibitem{14} C. Parrinello \& G. Jona-Lasinio, Phys. Lett. {\bf B251} (1990),
175. 
\bibitem{15} V.N. Gribov, Nucl. Phys. {\bf B139} (1978), 1. 
\bibitem{16} H. Verschelde, K. Knecht, K. van Acoleyen \& M. Vanderkelen, Phys.
Lett. {\bf B516} (2001), 307.
\bibitem{17} R.E. Browne \& J.A. Gracey, JHEP {\bf 11} (2003), 029.
\bibitem{18} H. Verschelde, Phys. Lett. {\bf B351} (1995), 242.
\bibitem{19} H. Verschelde, S. Schelstraete \& M. Vanderkelen, Z. Phys.
{\bf C76} (1997), 161.
\bibitem{20} T. van Ritbergen, A.N. Schellekens \& J.A.M. Vermaseren, Int. J.
Mod. Phys. {\bf A14} (1999), 41. 
\bibitem{21} S.G. Gorishny, S.A. Larin, L.R. Surguladze \& F.K. Tkachov,
Comput. Phys. Commun. {\bf 55} (1989), 381. 
\bibitem{22} S.A. Larin, F.V. Tkachov \& J.A.M. Vermaseren, ``The Form version
of Mincer'', NIKHEF-H-91-18. 
\bibitem{23} J.A.M. Vermaseren, math-ph/0010025.
\bibitem{24} D.R.T. Jones, Nucl. Phys. {\bf B75} (1974), 531; 
W.E. Caswell, Phys. Rev. Lett. {\bf 33} (1974), 244. 
\bibitem{25} O.V. Tarasov, A.A. Vladimirov \& A.Yu. Zharkov, Phys. Lett. 
{\bf B93} (1980) 429.
\bibitem{26} P. Nogueira, J. Comput. Phys. {\bf 105} (1993), 279. 
\bibitem{27} J.A. Gracey, Phys. Lett. {\bf B552} (2003), 101. 
\bibitem{28} S.A. Larin \& J.A.M. Vermaseren, Phys. Lett. {\bf B303} (1993), 
334. 
\end{thebibliography}
\end{document}